\pdfoutput=1
\documentclass[sort&compress]{elsarticle}
\usepackage{amssymb,amsmath}
\usepackage{units}

\usepackage{hyperref}

\def\l{\left(}
\def\r{\right)}

\newcommand{\be}{\begin{equation}}
\newcommand{\ee}{\end{equation}}
\newcommand{\ba}{\begin{align}}
\newcommand{\ea}{\end{align}}
\newcommand{\bg}{\begin{gather}}
\newcommand{\eg}{\end{gather}}
\newcommand{\bseq}{\begin{subequations}}
\newcommand{\eseq}{\end{subequations}}

\newcommand{\matm}{m_{\text{atm}}}%

\newcommand{\MeV}{\text{\,MeV}}
\newcommand{\GeV}{\text{\,GeV}}

\newcommand{\Br}{\mathop\mathrm{Br}}

\begin{document}
\title{Testing $\nu$MSM 
with indirect searches}

\author[inr,mipt]{Dmitry Gorbunov}
\ead{gorby@ms2.inr.ac.ru}
\author[inr,msu]{Inar Timiryasov}
\ead{timirysov@ms2.inr.ac.ru}

\address[inr]{Institute for Nuclear Research of the Russian Academy of
  Sciences,\\
  60th October Anniversary prospect 7a, Moscow 117312, Russia}
\address[mipt]{Moscow Institute of Physics and Technology, Dolgoprudny
  141700, Russia}
\address[msu]{Physics Department, Moscow State University, Vorobievy Gory,  
Moscow 119991, Russia}

\begin{abstract}
We consider neutrino Minimal extension of the Standard Model
($\nu$MSM), which by introducing only three sterile neutrinos in 
sub-electroweak region can explain active neutrino oscillations (via
seesaw type-I mechanism), baryon asymmetry of the Universe 
(leptogenesis via oscillations) and dark matter phenomena 
(with keV-scale sterile neutrino forming dark matter). 
We estimate sterile neutrino virtual
contributions to various lepton flavor and lepton number violating
processes. The contributions are too small, giving no chance for
indirect searches to compete with direct measurements in exploring
$\nu$MSM.
\end{abstract}


\maketitle


\paragraph{Introduction}
The Standard Model of elementary particle physics (SM) is an extremely successful
model, which has passed countless precision tests, and whose main predictions
have been confirmed.  
Its last but crucial missing ingredient,
required for the theory to be consistent --- the Higgs boson --- was discovered
at the LHC in 2012 \cite{Aad:2012tfa,Chatrchyan:2012ufa}.  

The \emph{neutrino flavor oscillations} --- transitions between neutrinos of
different flavors (see e.g.~\cite{Strumia:2006db} review) --- are among the few
firmly established phenomena \emph{beyond the Standard Model}. 
The direct coupling of the neutrino species (in the
form $\bar \nu_\alpha \nu_\beta$)\footnote{We use denotations where $\nu_\alpha
  = \{\nu_e,\nu_\mu,\nu_\tau\}$ are neutrinos interacting with
  $W$-boson, i.e. weak charge eigenstates.} %
is prohibited by the SU(2) gauge symmetry. The
oscillation phenomena can be described by a non-renormalizable operator of
``dimension 5'':
\begin{equation}
  \label{eq:1}
  \mathcal{L}_\text{osc} = \mathcal{L}_\textsc{sm} + c_{\alpha\beta}
  \,\frac{(\bar L_\alpha
    \cdot \tilde\Phi) (L_\beta^C
    \cdot \tilde\Phi^*) }{\Lambda},
\end{equation}
where $L_\alpha$ are left SU(2) doublets of leptons of different
flavors, $\alpha=e,\mu,\tau$, subscript $C$ refers
to the charge conjugation, $L_\alpha^C = i\sigma^2 L_\alpha^*$ and 
$\Phi$ is the Higgs SU(2) doublet, $\tilde \Phi_i = \epsilon_{ij}
\Phi^*_j$; $c_{\alpha\beta}$ is a dimensionless $3\times 3$ matrix. If
some of its elements are $\mathcal{O}(1)$, the scale of new physics
where the interaction \eqref{eq:1} must be replaced with a
renormalizable model is $\Lambda \sim
v^2/\matm \sim 10^{15}$~GeV, where $v = \sqrt{2} \langle \Phi\rangle =
\unit[246]{GeV}$.

The neutrino mixing term~(\ref{eq:1}) can be mediated via exchange of some new
particles (for example: gauge-singlet fermions; a scalar in an adjoint
representation of the SU(2) gauge group; a fermion in the adjoint
representation of the SU(2);
etc., see\,\cite{Strumia:2006db} for a review) interacting with left lepton
doublet $L_\alpha$ and with the Higgs doublet $\Phi$.  A traditional
explanation of the smallness of neutrino masses ($\matm \ll
m_\text{electron}$) is provided by the \emph{seesaw
  mechanism}. The small number is the ratio of the Dirac
mass of the neutrino to the mass of the new particle. 
The seesaw mechanism does not predict, however, the mass
of the new particles that can have any value.

If the new particles carry quantum number of the SU(2) gauge group, their
non-detection means that they should be heavier than the reach of modern
accelerators. The situation is different, however, if the term~(\ref{eq:1}) is
mediated via exchange of the gauge-singlet fermion (the so-called
\emph{type-I seesaw mechanism}~
 \cite{Minkowski:1977sc,Ramond:1979py,Mohapatra:1979ia,Yanagida:1980xy,Schechter:1980gr}, 
  see~\cite{Orloff:2005nu} and the references therein). In this case, the SM 
Lagrangian $\mathcal{L}_\textsc{sm}$ is extended by introducing $\mathcal{N}$
right-handed fermions $N_I$, $I=1,\dots,\mathcal{N}$:
\begin{equation}
\label{eq:2}
\mathcal{L} = \mathcal{L}_\textsc{sm} + i \bar{N}_I \gamma^\mu\partial_\mu N_I-
\left (F_{\alpha I} \bar{L}_\alpha N_I \tilde{\Phi} + \frac{M_I}{2}\bar{N}^C_I N_I +h.c. \right),
\end{equation}
where $F_{\alpha I}$ are new Yukawa couplings.  The Yukawa
interaction terms dictate the SM charges of the right-handed
particles: they turn out to carry no electric, weak and strong
charges; therefore they are often termed ``singlet fermions'' or
``sterile neutrinos''.  Sterile neutrinos can thus have Majorana
masses, $M_I$, consistent with the gauge symmetries of the SM. 
The number of these singlet fermions must be $\mathcal{N}\ge 2$
to explain the data on ``active'' neutrino oscillations.  In the case
of $\mathcal{N}=2$ there are \emph{11 new parameters} in the
Lagrangian~(\ref{eq:2}), while the neutrino mass/mixing matrix
$\mathcal{M}_\nu$ has 7 parameters in this case (including two active
neutrino masses, one of the three active neutrinos remains massless).
The situation is even more relaxed for ${\cal N}>2$.

Generically the remaining free parameters $F_{\alpha I}$ may take any
values (consistent with perturbativity), including those which make
sterile neutrinos even more valuable in particle physics. Indeed, the
lightest sterile neutrino can serve as a dark matter candidate, while
violating lepton number Majorana mass term and CP-violating complex
phases of Yukawa matrix in \eqref{eq:2} provide necessary
conditions for leptogenesis via active-sterile neutrino oscillations
in the early Universe. If happened before the electroweak phase
transition, it can yield the baryon asymmetry of the plasma thus
explaining the matter-antimatter asymmetry in the later
Universe. Successful solution of both problems (dark matter and baryon
asymmetry of the Universe) requires ${\cal N}\geq3$ sterile
neutrinos. 

\paragraph{$\nu$MSM: description and phenomenology} 
The most economic model of this type, which recruits
${\cal N}=3$ sterile neutrinos, is known as
$\nu$MSM \cite{Asaka:2005an,Asaka:2005pn} 
that is \emph{neutrino Minimal extension of the Standard Model}. 
It turns out that sterile neutrino dark matter particle mass is
confined in $M_1\sim 1-50$\,keV region, 
see \cite{Boyarsky:2009ix} for review. The requirement of dark matter
stability on cosmological timescales makes its coupling with the SM 
species so feeble, that it does not contribute significantly to the
neutrino oscillation pattern~\cite{Boyarsky:2006jm}. In this sense, the
$\nu$MSM setup is close to that of ${\cal N}=2$ seesaw scheme: 
one of the active neutrino is (almost) massless. The two heavier
sterile neutrinos are responsible for both active neutrino masses and
baryon asymmetry of the Universe (BAU). Actually, to produce enough lepton
asymmetry before the electroweak phase transition and yet not
equilibrate sterile neutrino in the primordial plasma one must resort
to a resonance enhancement of active-sterile oscillations in the early
Universe \cite{Akhmedov:1998qx,Asaka:2005pn}, 
achieved with almost degenerate neutrinos, 
\begin{equation}
\label{degeneracy}
\left| M_3-M_2\right | \equiv \Delta M\ll M_{2,3}\,,
\end{equation}
that may or may not originate in a (slightly broken) global $U(1)$
symmetry\,\cite{Shaposhnikov:2006nn}. 
Moreover, the particular mechanism of the leptogenesis works for
sterile neutrinos below the electroweak scale, though an exact upper
limit on the sterile neutrino mass has not been settled yet. 

\begin{figure}[!htb]
  \centerline{
  \includegraphics[width=0.45\textwidth]{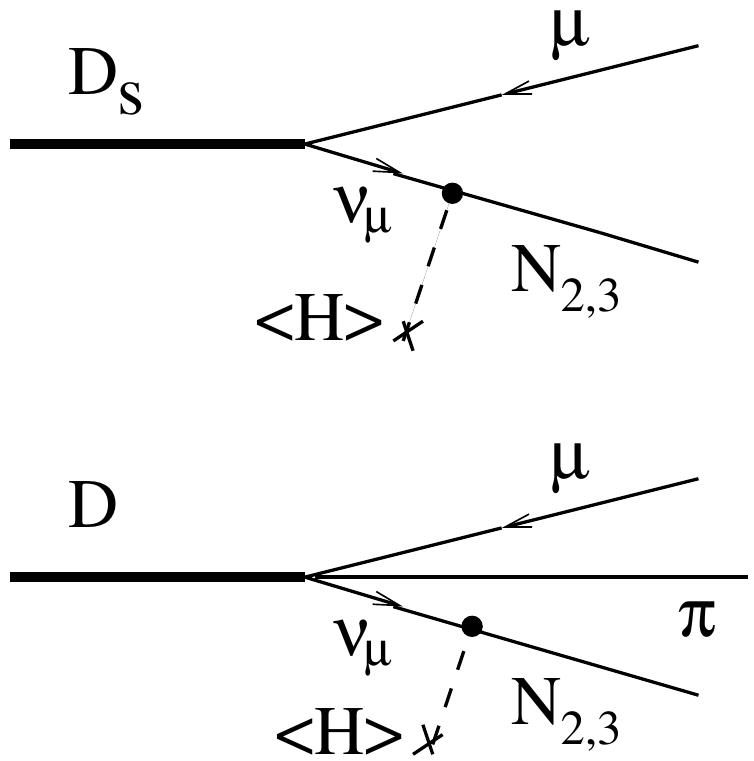}
\hskip 0.1\textwidth
\includegraphics[width=0.45\textwidth]{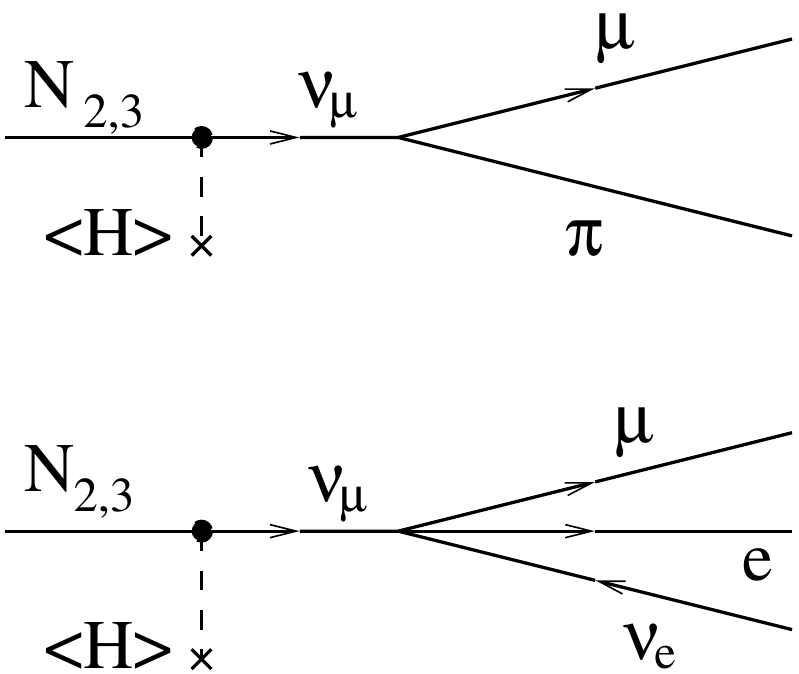}}
  \caption{Typical diagrams of sterile neutrino production {\it (left
  panel)} and decays {\it (right panel)}.}
  \label{production-and-decays}
\end{figure}

In all the relevant dynamics the active-sterile mixing rising from
Yukawa terms in \eqref{eq:2} plays the major role.  Fermions $N_I$
in \eqref{eq:2} are \emph{charge eigenstates} of weak interactions and
they are truly neutral. However, due to the Yukawa mixing the mass
eigenstates have small admixture of $\nu_\alpha$ characterized by 
small active-sterile mixing angle
\begin{equation}
\label{mixing-angle}
U_{\alpha I}\equiv\frac{v\, F_{\alpha I}}{\sqrt{2}\,M_I}\;,
\end{equation}
$|U_{\alpha I}|\ll 1$, 
and as a result carry small weak
charge. For a given sterile neutrino 
its smallness is characterized by the following dimensionless number:
\begin{equation}
  \label{eq:3}
U^2_I \equiv\sum_\alpha \frac{v^2 |F_{\alpha I}|^2}{2\,M_I^2}\,.
\end{equation}
This means that from a phenomenological point of view, particles $N_I$ behave
like \emph{heavy neutral leptons}. 
In what follows, we denote the heavy mass eigenstates in the same way
as the charge eigenstates, $N_I$, because in $\nu$MSM the admixture of
$\nu_\alpha$ is very small indeed.
These particles can be created in decays of
other particles instead of the usual neutrino $\nu_\alpha$ (if kinematics
allows it), and in turn decay into the SM particles, 
see e.g.\ Fig.~\ref{production-and-decays}. 
Both production and decay
rates are proportional to the squared mixing
angles \eqref{mixing-angle}: the smaller the mixing, the lower the
rates.  Based on this phenomenology, a number of searches of heavy neutral
leptons have been performed in the past, yielding exclusion regions in
the $(U_{\alpha I},M_I)$ parameter space, for review, see e.g. \cite{Atre:2009rg}.

\begin{figure}[!htb]
\centerline{
\includegraphics[width=0.5\textwidth]{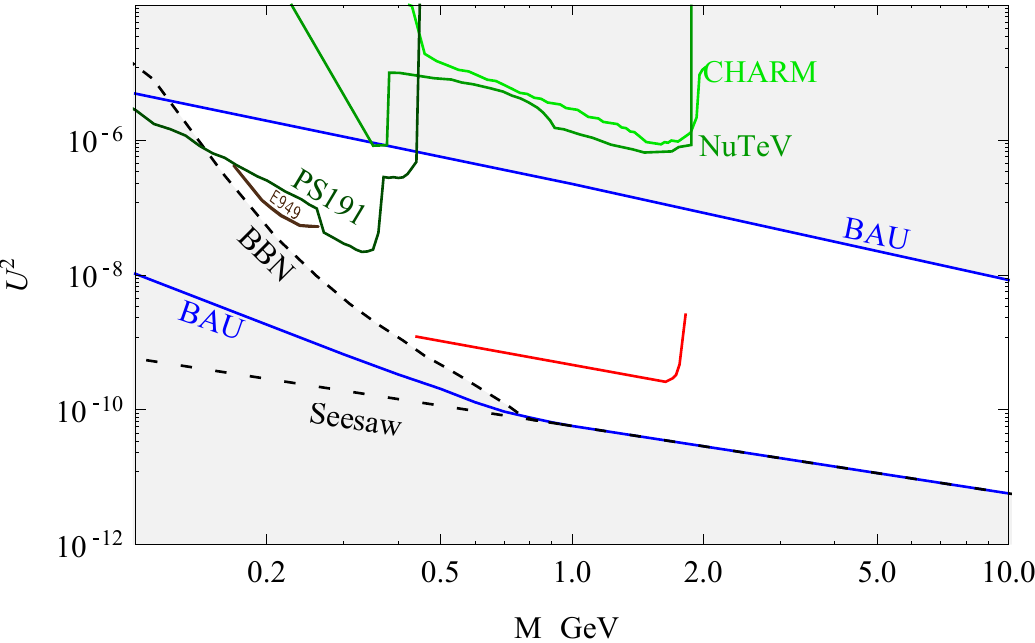}
\includegraphics[width=0.5\textwidth]{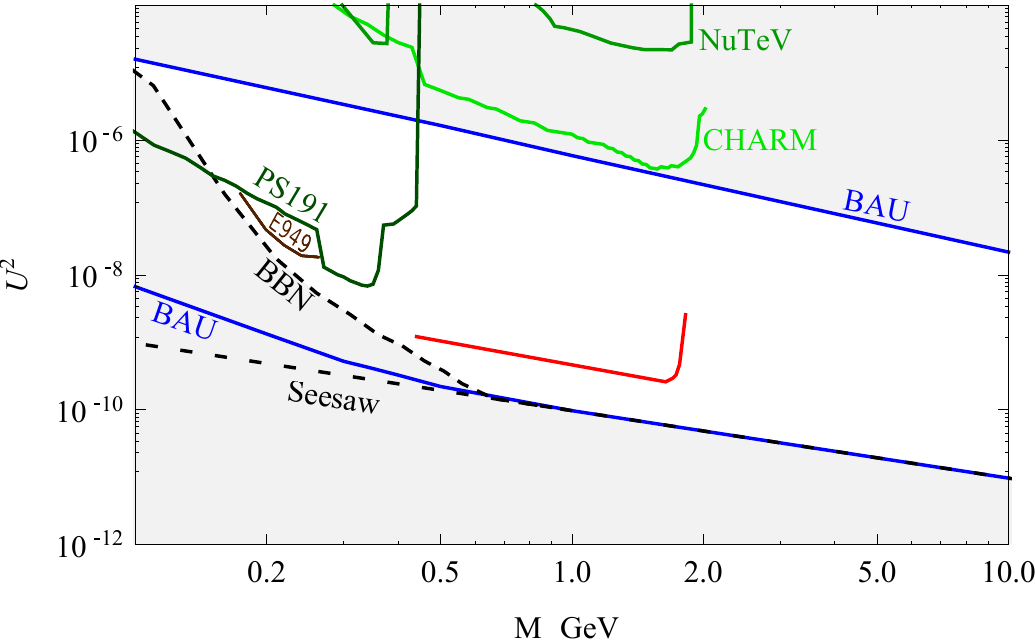}
}
\caption{Viable region of $\nu$MSM parameter space for the cases of
normal {\it (left panel)} and inverted {\it (right panel)} hierarchies
in active neutrino sector, adopted from \cite{Gninenko:2013tk}.  
Upper limits are from successful baryogenesis
(BAU) and from direct searches (PS191, CHARM, NuTeV, 
and very recent E949\,\cite{Artamonov:2014urb}), lower limits
are from seesaw mechanism (seesaw), from Big Bang Nucleosynthesis (BBN)
and from successful baryogenesis (BAU). There is also an unnamed line
indicating a possible sensitivity of a dedicated fixed target
experiment proposed in \cite{Gninenko:2013tk}.     
\label{parameter-space}
}
\end{figure}

In $\nu$MSM with almost decoupled dark matter neutrino and 
degenerate two heavy sterile neutrinos the direct searches place 
limits on the flavor mixing angles  
\begin{equation}
  \label{eq:5}
  U_\alpha^2 \equiv 
  \sum_{I=2,3} \frac{v^2 |F_{\alpha I}|^2}{2\,M_I^2} \;,
\end{equation}
while cosmology constrains mostly 
the sum of the angles~(\ref{eq:5}) over all relevant flavors:
\begin{equation}
\label{eq:4}
U^2\equiv\sum_{I=2,3}U^2_I = \sum_{\alpha=e,\mu,\tau} U^2_\alpha\,,
\end{equation}
that is the overall mixing strength. Sterile neutrinos produced in
the early Universe can decay and destroy the primordial chemical
elements.  
For successful Big Bang Nucleosynthesis (BBN), the lifetime $\tau_N$ of heavy
sterile neutrinos $N_{2,3}$ 
is restricted to be shorter than $0.1$\,s\,\cite{Dolgov:2000pj}.  
This restriction yields a 
lower bound on overall mixing $U^2$ \cite{Gorbunov:2007ak}. 
Then, successful generation of the lepton asymmetry asks for the
sterile neutrinos to be out-of-equilibrium in the plasma, which
places an upper limit (with a factor of few uncertainty) 
on the mixing \cite{Canetti:2012vf}. A simple numerical fit to presented
in Ref.\,\cite{Canetti:2012vf} results reads  
\begin{equation}
\label{seesaw-limit}
U^2<\kappa \times 2.5\times 10^{-7} \l \frac{\GeV}{M_N}\r^{3/2}\;, 
\end{equation}
where $\kappa=1(2)$ refers to the normal (inverted)
hierarchy in active neutrino sector. Leptogenesis gives also
lower limits on mixing referring to the minimal connection between
active and sterile neutrino sectors which still provides enough
baryon asymmetry in the end.  Finally, a lower bound on
mixing \cite{Gorbunov:2007ak,Canetti:2012vf} 
\begin{equation}
\label{BAU-limit}
U^2>5\,\kappa\times 10^{-11}\frac{\GeV}{M_N}
\end{equation}
is inherited in the seesaw mechanism: at a given sterile neutrino
masses the mixing can not be arbitrary small, as it determines 
the active neutrino masses, and two of them must exceed 0.05\,eV and
0.008\,eV to be consistent with atmospheric and solar neutrino 
oscillation data \cite{Agashe:2014kda}, respectively. The
aforementioned constraints are presented in Fig.\,\ref{parameter-space} 
There are also limits on the model parameter space associated with
dark matter, however, they are largely depend on the mechanism of dark
matter production in the early Universe (for working examples and
discussion, see 
e.g.\,\cite{Shaposhnikov:2006xi,Bezrukov:2009yw,Canetti:2012vf,Canetti:2012kh,Bezrukov:2014nza}),  
hence we ignore them for the sake of generality.

The best way to explore the viable region of model parameter space
outlined in Fig.\,\ref{parameter-space} is direct searches for sterile
neutrino production and decays, see \cite{Gorbunov:2007ak} for
details. For light neutrino masses, $M<5$\,GeV, these searches can be performed
with a fixed target experiment operating on high energy and high
intensity proton beam, which produces heavy hadrons on target and the
latter decay into sterile neutrinos as shown in
Fig.\,\ref{production-and-decays}. The suitable experimental setup was
sketched in Ref.\,\cite{Gninenko:2013tk} and evolved to the proposal
of a new beam dump experiment at CERN \cite{Bonivento:2013jag} later
called SHiP. For heavier mass the only option is high-energy
colliders, and $e^+e^-$ machines producing many billions of
$Z$-bosons (which immediately decaying into sterile neutrinos) are
advocated \cite{Blondel:2014bra} to be very sensitive to $\nu$MSM. 

At the same time, $\nu$MSM  can be tested indirectly due to virtual
sterile neutrino contribution to particular processes with lepton
number or lepton flavor violation. In this {\it Letter} we
obtain the $\nu$MSM predictions for the bunch of relevant 
processes. As we found, the expected rates, except neutrinoless double
$\beta$-decay\footnote{Which is almost always (except narrow mass
interval $M_N\approx 150-250$\,MeV) saturated by two active massive
neutrinos, and hence insensitive to the model.},  
are so tiny, that they leave no chance at all to 
find any hint of $\nu$MSM physics indirectly. These findings confirm
that the direct searches are not only superior but the only realistic
way to explore the model.

Even with the seesaw, BBN and BAU constraints satisfied, 
there is a lot of freedom in relations between different Yukawa
couplings $F_{\alpha I}$. If sterile neutrino mass degeneracy follows
from the broken $U(1)$ symmetry, their mixing to the active neutrinos
are naturally degenerate as
well \cite{Shaposhnikov:2006nn,Gorbunov:2007ak}, so $\left |U_{\alpha
2}\right|^2\approx\left|U_{\alpha 3}\right|$. 
To present quantitative predictions three
different sets of couplings were considered in
\cite{Gorbunov:2007ak}. These ``extreme models''  are formulated in such
a way that coupling to a single active neutrino flavor dominates:
\begin{align*}
{\rm model~I}\;:& \hskip 0.5cm |U_{eI}|^2:|U_{\mu I}|^2:|U_{\tau
I}|^2 \approx 52:1:1\;,
~~~\kappa=2\;,\\
{\rm model~II}\;:& \hskip 0.5cm |U_{eI}|^2:|U_{\mu I}|^2:|U_{\tau
I}|^2 \approx 1:16:3.8\;,
~~~\kappa=1\;,\\
{\rm model~III}\;:& \hskip 0.5cm |U_{eI}|^2:|U_{\mu I}|^2:|U_{\tau
I}|^2 \approx 0.061:1:4.3\;,
~~~\kappa=1\;.
\end{align*}

Here and below dark matter $N_1$ is ignored and index
$I$ runs through 2 and 3 only. 

Recall that with effective framework of 2 sterile neutrinos relevant
for $\nu$MSM the 
type-I seesaw model brings 11 extra parameters while neutrino
oscillation data can fix at most 7 (one eigenstate is 
massless then). As a result, 4 parameters remain free, including one
CP-violating phase in the active-sterile Yukawa matrix. 
For some special values of this phase accidental cancellations of
sterile neutrino contribution in certain processes are possible, see
examples
in \cite{Shaposhnikov:2006nn,Asaka:2011pb,Ruchayskiy:2011aa}. 
We did not consider this effect hereafter, as it can only make the effects of
sterile neutrinos weaker.

\vskip 0.3cm 
We start with \emph{lepton flavor} violating processes. 

\paragraph{$l\to l'\gamma$}  
Decay rate of $l\to l'\gamma$ is given by \cite{Cheng:1980tp,Ilakovac:1994kj}:
\begin{equation}
\Gamma(l\to l' \gamma)=\frac{3\alpha}{8\pi}\frac{G_F^2m_l^5}{192\,\pi^3}\left 
|\sum_I U_{eI}U^*_{\mu I}\, g\l\frac{M_{N_I}^2}{m_W^2}\r\right |^2\!,
\label{Brmue}
\end{equation}
where $G_F$ is the Fermi constant, $\alpha$ is the fine structure
constant, $m_W$ is the mass of $W$-boson and 
\begin{equation*}
g(x)=\frac{x(1-6x+3x^2+2x^3-6x^2 \log x)}{2(1-x)^4}\,.
\end{equation*}

\begin{figure}[!htb]
\centerline{
\includegraphics[width=0.485\textwidth]{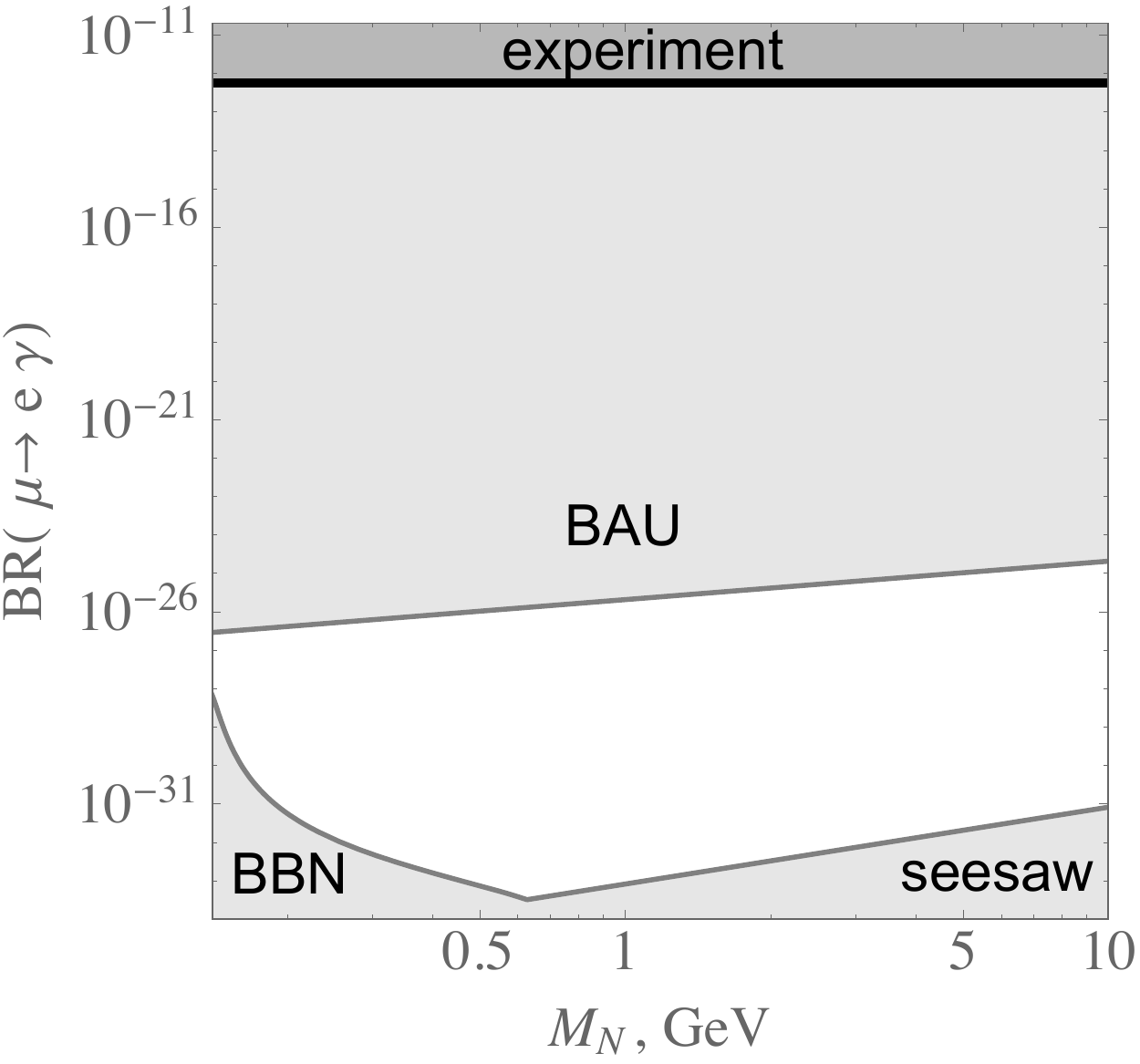}
\hskip 0.03\textwidth 
\includegraphics[width=0.485\textwidth]{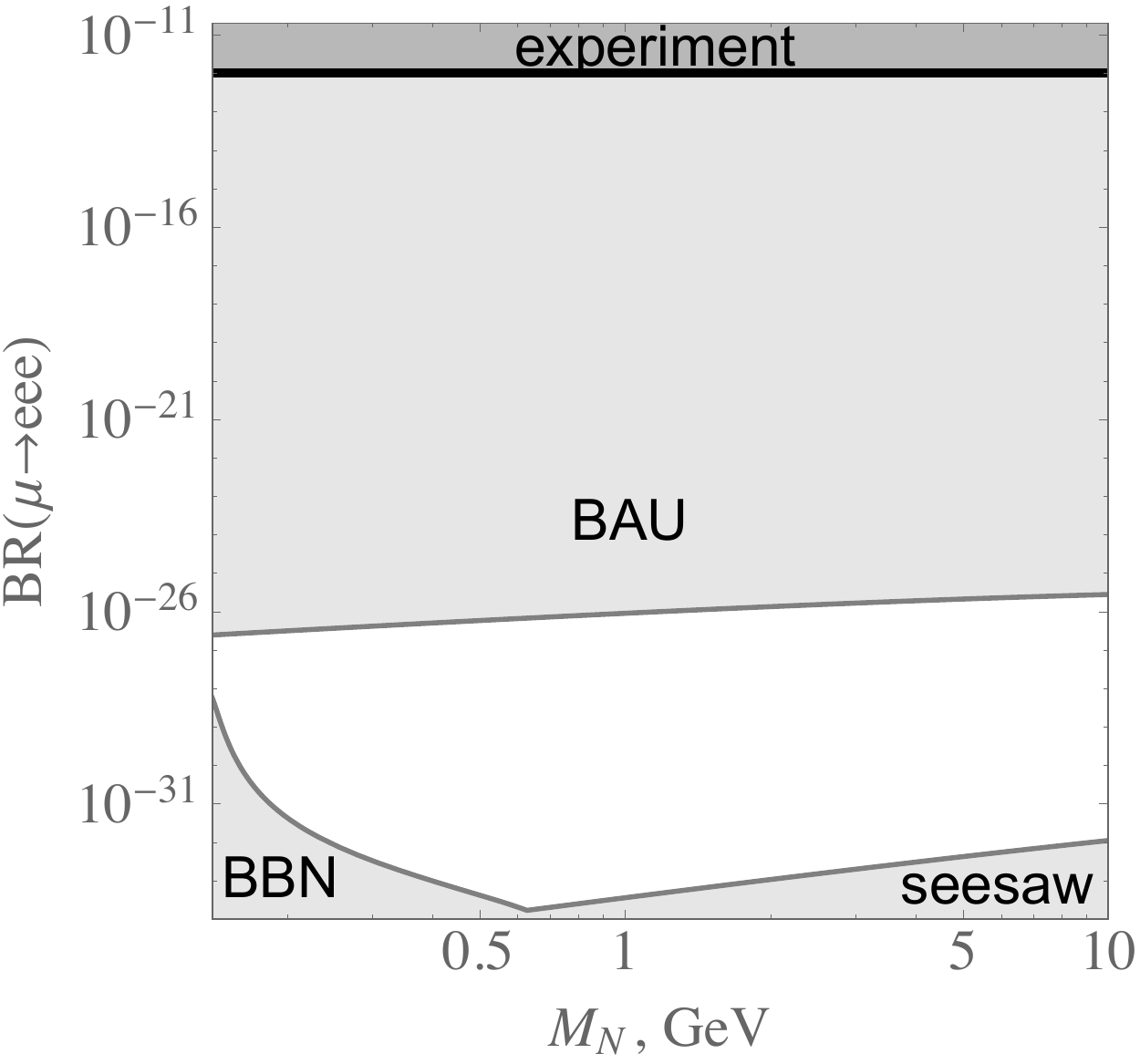}}

\vskip 0.4cm
\centerline{\includegraphics[width=0.485\textwidth]{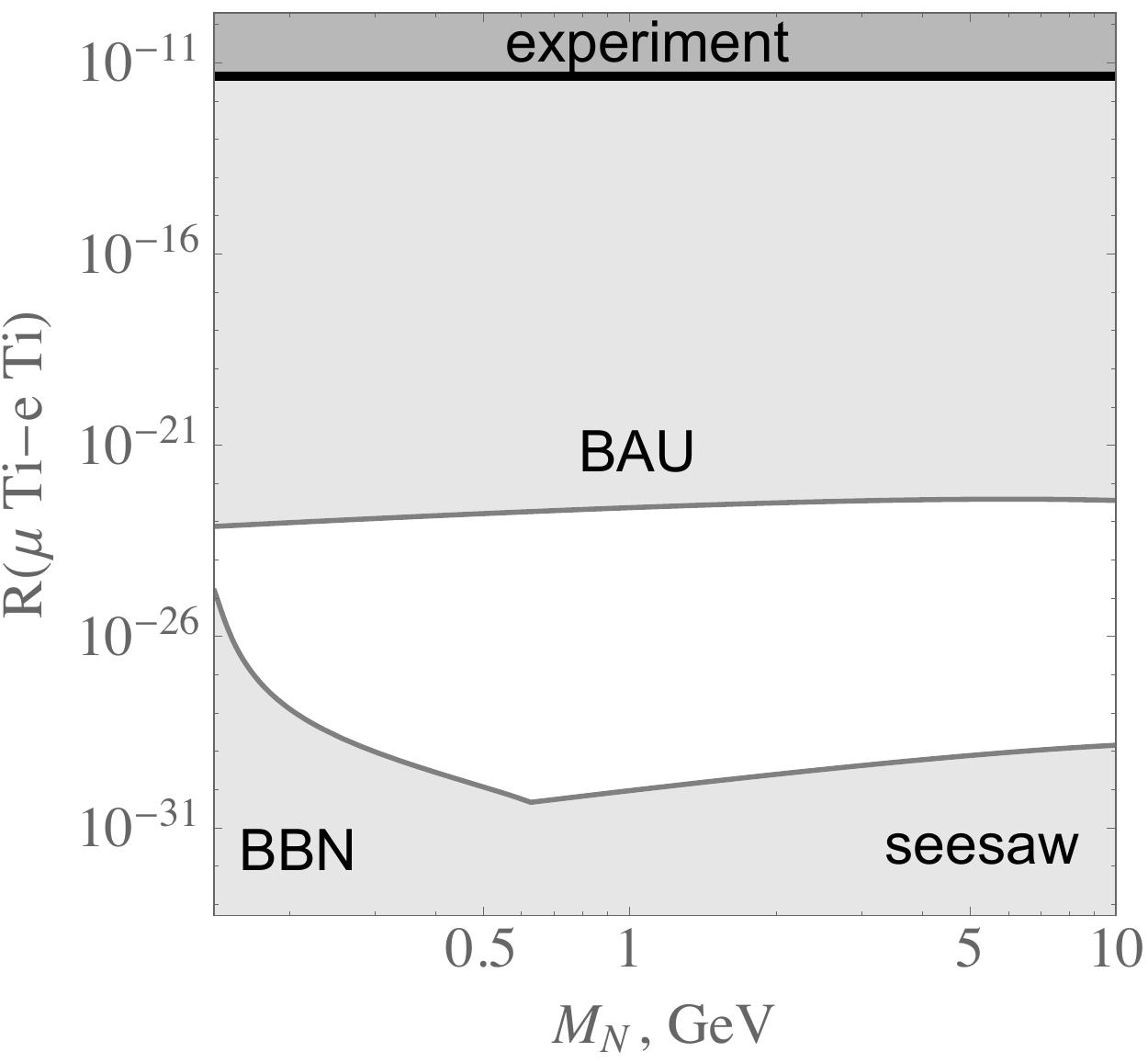} 
\hskip 0.03\textwidth
\includegraphics[width=0.485\textwidth]{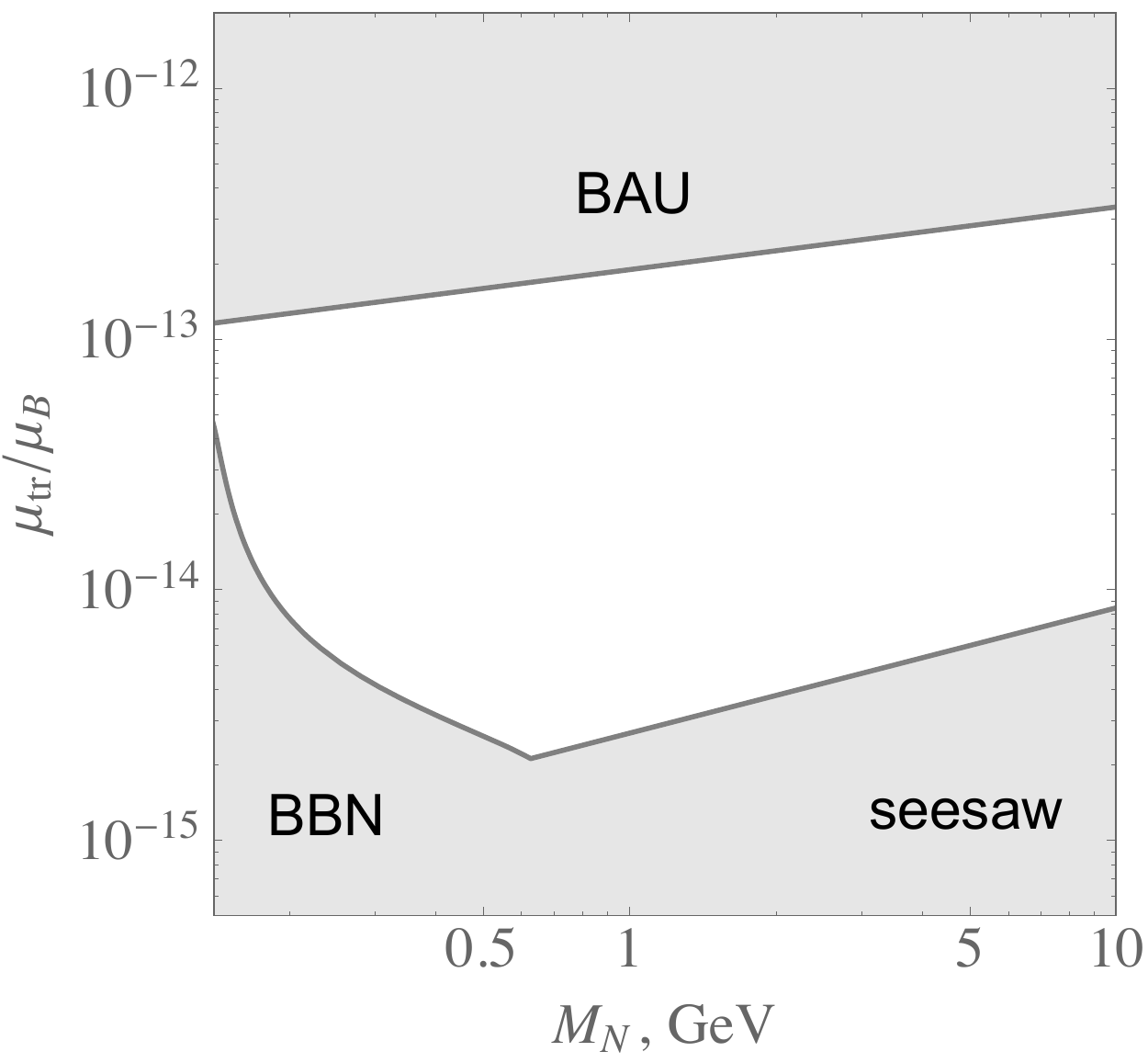}}
\caption{Decay branching ratios Br$(\mu\to e\gamma)$ {\it (top left panel)}, 
Br$(\mu\to eee)$ {\it (top right panel)}, conversion rate  
$R(\mu Ti\to e Ti)$ {\it (bottom left panel)} and neutrino transition dipole
moment $\mu_{\text{tr}_{\alpha I}}$ {\it (bottom right panel)} 
as functions of the sterile neutrino mass $M_N$ in the model I
(which exhibits the highest rate): 
white region is allowed. The region above solid line is 
excluded by experimental searches.}
\label{mueee}
\end{figure}

Inserting the limits on $U_{\alpha I}(M_N)$
discussed above \eqref{seesaw-limit}, \eqref{BAU-limit} 
 into eq.\,\eqref{Brmue} 
one can compare the expected in $\nu$MSM 
branching ratios of $l\to l'\gamma$ decays 
with existing experimental limits. We present
the result for $\mu\to e\gamma$ on the left top panel in Fig.\,\ref{mueee}.
Solid line shows the experimental limit 
\cite{Agashe:2014kda} $\Br(\mu\to e \gamma)_{\text{exp}}<5.7\times 10^{-13}$.
One can see that it is much less stringent than
cosmological and seesaw constraints. 
Experimental limits for two other relevant processes
are even weaker \cite{Agashe:2014kda}, $\Br(\tau\to e
\gamma)_{\text{exp}}<3.3\times 10^{-8}$ and $ \Br(\tau\to \mu
\gamma)_{\text{exp}}<4.4\times 10^{-8}$, while the $\nu$MSM predictions are
lower.

\paragraph{$\mu-e$ conversion} 
Another potentially interesting process is $\mu - e$ conversion in
nuclei. As was shown in \cite{Alonso:2012ji} for the quasi-degenerate
spectrum of sterile neutrinos, which is the case for $\nu$MSM, the
ratio
\begin{equation}
R^{\mu-e}_{\mu\to e\gamma}=\frac{R(\mu Z-e Z)}{Br(\mu\to e \gamma)},
\end{equation}
where $R(\mu Z-e Z)$ is the conversion rate in the nuclei of electric 
charge $Z$, could be sufficiently large, $\sim 10^3$, 
for light, $M_N<10\GeV$, sterile neutrinos.  This fact 
and expecting experimental progress \cite{Hungerford:2009zz} make 
$\mu - e$ conversion much more promising in testing $\nu$MSM as
compared to other lepton flavor violating processes. However, the
$\nu$MSM predictions are still some 5 orders of magnitude  
below the sensitivity of future experiments, that is in agreement with
Ref.\,\cite{Canetti:2013qna}. The results based on calculations
performed in Ref.\,\cite{Alonso:2012ji} are presented
on left bottom panel in Fig.\,\ref{mueee}.

\paragraph{$l\to l'l'l''$ } 
Branching ratio of the process $l\to l'l'l''$ was calculated in
\cite{Ilakovac:1994kj}. 
Present experimental limit on the branching ratio of 
$\mu^-\to e^-e^+e^-$ is \cite{Agashe:2014kda} $\Br(\mu^-\to
e^-e^+e^-)_{\text{exp}}<1.0\times 10^{-12}$.  Predictions of $\nu$MSM for the branching 
ratio of this process are shown on right top panel in Fig.\,\ref{mueee}. 
Present experimental limits on similar $\tau$-lepton decays
are typically weaker by four orders of magnitude (i.e. $\Br(\tau^-\to
e^-\mu^+e^-)_{\text{exp}}<1.5\times 10^{-8}$), while $\nu$MSM
predictions are lower, than for the muon decay. 

\vskip 0.3cm 
We proceed with \emph{lepton number} violating processes relevant for
probing $\nu$MSM. There are three related
types of these processes: neutrinoless double beta decay, decays of
$\tau$-lepton and meson decays into the same-sign leptons.

\paragraph{Neutrinoless double beta decay} 
The process of neutrinoless double beta decay is on the keen interest
both from the experimental and theoretical points of view since 
it is naturally dominated by active neutrinos. The 
effective mass $m_{\beta\beta}$ standing in front of the decay
amplitude depends on 7 out of the 9 parameters of active neutrino
sector: 
\begin{equation}
m^\nu_{\beta\beta} =| \sum_{i}m_i U_{e i}^2 |,
\label{mbb}
\end{equation}
where $m_i$ are active neutrino masses and $U_{e i}$ are elements of 
the Pontecorvo--Maki--Nakagawa--Sakata matrix describing active neutrino 
mixing. As we have already explained,   
in the $\nu$MSM model the lightest active neutrino is 
nearly massless: $m_1<10^{-5}$\,eV \cite{Asaka:2005an}.
This fact leads to specific bounds on the effective mass since
existing limits depend on the mass of the lightest neutrino
 \cite{Rodejohann:2012xd}. 

\begin{figure}[!htb]
\centerline{
\includegraphics[width=0.485\textwidth]{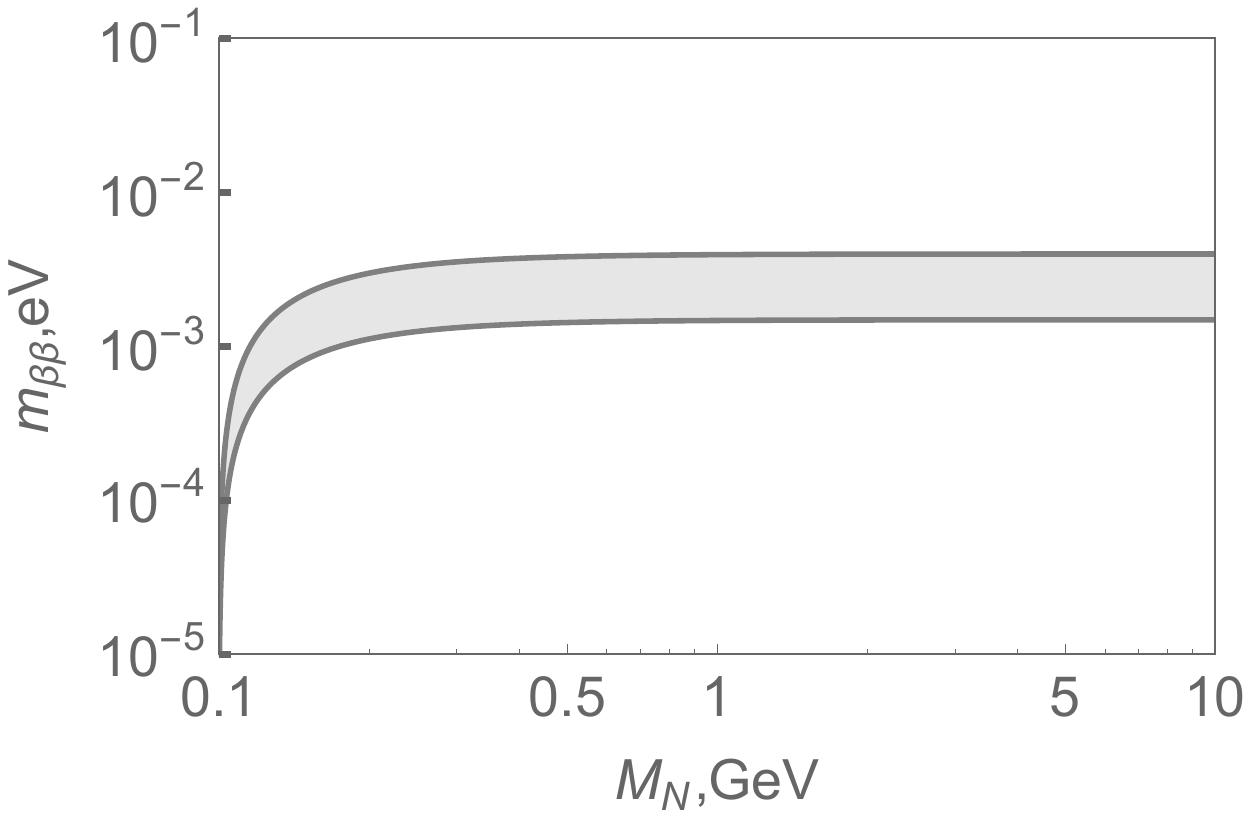}
\hskip 0.03\textwidth 
\includegraphics[width=0.485\textwidth]{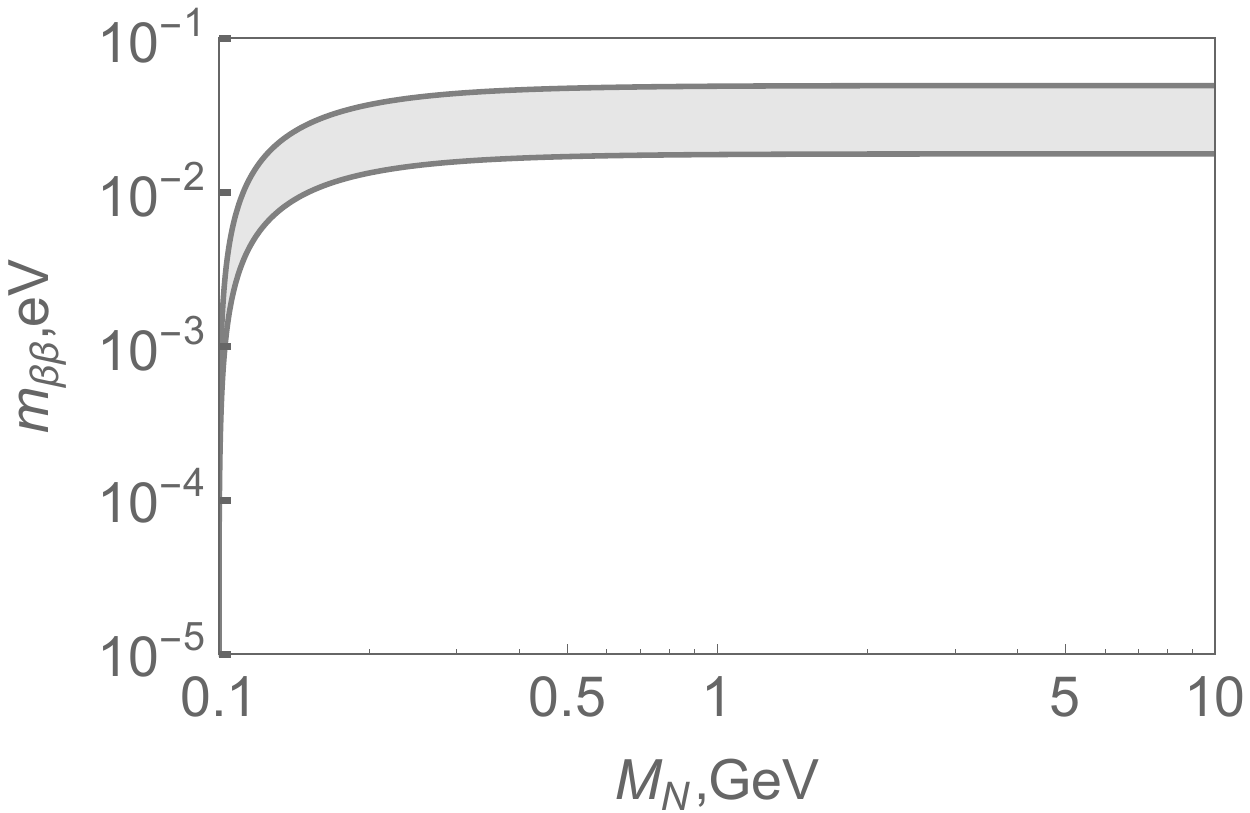}
}
\caption{Shadowed regions show $m_{\beta\beta}$ as function of $M_N$ for normal {\it (left
panel)}  
and inverted {\it (right plot)} hierarchies obtained with scan over
the unknown CP-phase in active neutrino sector. Asymptotics at large
masses agree with previous studies \cite{Bezrukov:2005mx}
if active neutrino mixing parameters are set to the presently known values
\cite{Agashe:2014kda}.}
\label{fig:mbb}
\end{figure}

Sterile neutrinos themselves also contribute to \eqref{mbb}.  As 
shown in \cite{Asaka:2011pb} their contribution to the effective neutrino
mass is destructive:
\begin{equation}
\label{mass2beta}
m_{\beta\beta}=[1-f(M_N)] m^\nu_{\beta\beta}\,,
\end{equation}
where function $f(M_N)$ describes the resulting suppression of the
decay amplitude, which depends on the sterile neutrino mass. 
Following analysis in \cite{Asaka:2011pb} we assume that 
$f(M_N)=1$ for $M_N<\Lambda_\beta$
and for heavier sterile neutrinos it decreases as
$f(M_N)=(\Lambda_\beta/M_N)^2$, where the typical energy scale of the
process is $\Lambda_\beta=100\MeV$.  Constraints on the effective mass
$m_{\beta\beta}$ as function of the mass of heavier sterile neutrinos
are shown in Fig. \ref{fig:mbb}.
At large sterile neutrino masses and for the \emph{present central values of
the known active neutrino sector parameters} the allowed intervals in
cases of normal (NH) and inverted hierarchies (IH) are
\begin{equation}
 \label{2beta}
1.5\,\text{meV}<m_{\beta\beta}^{\text{NH}}<3.9\,\text{meV}\,,\;\;\;\;
17\,\text{meV}<m_{\beta\beta}^{\text{IH}}<49\,\text{meV}\,.
\end{equation}
If neutrinoless double beta decay will be found with the rate corresponding
to the effective mass above the upper bound in \eqref{2beta} it would
imply additional (to $\nu$MSM) new physics in the neutrino sector.  

\paragraph{Decays into the same sign pairs} 
Branching ratios for lepton number violating $\tau$-lepton decays
(e.g. $\Br(\tau^- \to e^+ \pi^- \pi^-)$) and for meson decays into the
same sign charged leptons (e.g. $K^- \to e^+ e^+ \pi^-$) can be
obtained by making use of formulas in Appendix B of
Ref.\,\cite{Atre:2009rg} 
\emph{without adopting the narrow width approximation for integration
over the phase space}\footnote{Results obtained within that
approximation, including many presented in Ref.\,\cite{Atre:2009rg},  
are often very confusing, as for a  
really narrow resonance (hence long lived) they are applicable only for 
describing of signal events in a very large detector, 
which size exceeds the resonance decay length.}: for sterile neutrino
heavier than decaying particles the approximation is not applicable,
and for lighter sterile neutrino it is produced in the decay and 
can subsequently decay within a detector, for $\nu$MSM both processes are 
described in \cite{Gorbunov:2007ak} and are naturally associated with 
 \emph{direct searches}.   Decay widths for these
processes with \emph{heavy neutrino} 
are proportional again to the $U^4$ and hence resulting
bounds on branching ratios are far beyond the grasp of future
experiments. To illustrate this statements we have
calculated the decay rate of 3-body $\tau$-lepton decay and found
numerically for the branching ratio 
\[
\Br(\tau^- \to e^+ \pi^- \pi^-)\simeq 6\times 10^{-31}\times 
\l \frac{U^2}{10^{-8}}\r^{\!\!2} \l\frac{5\,\GeV}{M_I}\r^{\!2}\;. 
\]

Quite the same situation is realized for other similar decay modes and
decaying mesons.  One concludes that the number of weakly
decaying heavy particles (leptons and mesons) collected to check $\nu$MSM
predictions must be unrealistically big.




\paragraph{Neutrino transition dipole moments} 
Heavy sterile neutrino can radiatively decay into active neutrino and
photon (one-loop level process). 
Radiative decay width reads\,\cite{Pal:1981rm}:
\begin{equation}
\label{radiative-decay}
\Gamma(N_I\to\gamma\nu_\alpha)=\frac{9\alpha
  G_F^2}{1024\pi^4}\sin^2(2U_{\alpha I})M_I^5
\approx 2.6\times10^{-16}\times U_{\alpha I}^2\l\frac{M_I}{\GeV}\r^5 
\GeV,
\end{equation}
Since both neutrinos are electrically neutral, this process can be
effectively described by the transition dipole moments
$\mu_{\text{tr}\,\alpha I}$, see\,\cite{Beg:1977xz}. 
From \eqref{radiative-decay} we obtain the following estimate:
\[
\mu_{\text{tr}\,{\alpha I}}\approx 2.7\times
10^{-10}\times \mu_{\text{B}}\times U_{\alpha I} \,\frac{M_I}{\GeV}
\]
where $\mu_{\text{B}}$ is the Bohr magneton. Transition dipole moments
contribute to neutrino scattering off nuclei and can be probed with
large neutrino flux. Since the sterile neutrino is not observed in
this type of searches, we consider it among \emph{indirect searches} for sterile
neutrinos. Given the smallness of mixing predicted in $\nu$MSM, see
eq.\,\eqref{BAU-limit}, 
the transition dipole moments are too small to be tested in future 
experiments. The corresponding results are presented in
Fig.\,\ref{mueee}.



\paragraph{Conclusions} 
To summarize, we have obtained $\nu$MSM predictions for lepton flavor
and lepton number violating processes and observed that they are far
below the expected sensitivity of present and foreseeable future
experiments. Thus to probe $\nu$MSM one must resort to direct searches
only, see \cite{Gorbunov:2007ak,Blondel:2014bra} for examples.

\vspace{0.5cm}
 
We thank F.\,Bezrukov, W.\,Bonivento, M.\,Shaposhnikov for valuable
comments and O.\,Ruchayskiy for participating in the project at
earlier stage.   
The work was supported by the RSCF grant 14-22-00161.


\bibliography{\jobname}
\bibliographystyle{elsarticle-num}

\end{document}